\documentclass[twocolumn]{aastex631}

\usepackage{graphicx} 
\usepackage{amssymb}
\usepackage{float}
\usepackage{hyperref}
\usepackage{natbib}
\usepackage{enumitem}
\usepackage{soul}
\setenumerate[1]{label=(\arabic*)}

\newcommand\gaia{{\em Gaia}}
\newcommand\teff{T$_{\rm eff}$}
\newcommand\logg{log~$g$}
\newcommand\solarmass{M$_\odot$}

\begin{document}

\title{Hunting for Polluted White Dwarfs and Other Treasures with \gaia~XP Spectra and Unsupervised Machine Learning}

\author[0000-0001-5745-3535]{Malia L. Kao}
\affiliation{Department of Astronomy, University of Texas at Austin, 2515 Speedway, Austin, TX 78712, USA}

\author[0000-0002-1423-2174]{Keith Hawkins}
\affiliation{Department of Astronomy, University of Texas at Austin, 2515 Speedway, Austin, TX 78712, USA}

\author[0000-0002-3553-9474]{Laura K. Rogers}
\affiliation{Institute of Astronomy, University of Cambridge, Madingley Road, Cambridge, CB3 0HA, UK}

\author{Amy Bonsor}
\affiliation{Institute of Astronomy, University of Cambridge, Madingley Road, Cambridge, CB3 0HA, UK}

\author[0000-0002-1086-8685]{Bart H. Dunlap}
\affiliation{Department of Astronomy, University of Texas at Austin, 2515 Speedway, Austin, TX 78712, USA}

\author[0000-0003-4593-6788]{Jason L. Sanders}
\affiliation{Department of Physics \& Astronomy, University College London, Gower Street, London, WC1E 6BT, UK}

\author[0000-0002-6748-1748]{M. H. Montgomery}
\affiliation{Department of Astronomy, University of Texas at Austin, 2515 Speedway, Austin, TX 78712, USA}

\author{D. E. Winget}
\affiliation{Department of Astronomy, University of Texas at Austin, 2515 Speedway, Austin, TX 78712, USA}

\begin{abstract}

White dwarfs (WDs) polluted by exoplanetary material provide the unprecedented opportunity to directly observe the interiors of exoplanets. However, spectroscopic surveys are often limited by brightness constraints, and WDs tend to be very faint, making detections of large populations of polluted WDs difficult. 
In this paper, we aim to increase considerably the number of WDs with multiple metals in their atmospheres. Using 96,134 WDs with \gaia~DR3 BP/RP (XP) spectra, we constructed a 2D map using an unsupervised machine learning technique called Uniform Manifold Approximation and Projection (UMAP) to organize the WDs into identifiable spectral regions. The polluted WDs are among the distinct spectral groups identified in our map. We have shown that this selection method could potentially increase the number of known WDs with 5 or more metal species in their atmospheres by an order of magnitude. Such systems are essential for characterizing exoplanet diversity and geology. 

\end{abstract}


\section{Introduction}

The death of a low-mass ($\lesssim$ 8 \solarmass) main-sequence star culminates in the ejection of its outer layers in a planetary nebula and the collapse of its core into a white dwarf (WD). Typical mass WDs are extremely dense and have very high surface gravities (\logg~$\sim$ 8.0, central density $\approx 10^6$ $\mathrm{g/cm^3}$), equivalent to about 100,000 times that of Earth. As such, WDs are chemically stratified, meaning lighter elements like H and He rise to the surface and heavier elements (e.g., C, O, Ca, Mg, Fe, etc.) sink to the core. The majority of WDs are expected to consist of carbon-oxygen cores and thin upper layers of He and H that make up only $\sim$ 1\% of the total WD mass. For most WD spectra, we expect the presence of H (DA) or He {\small I} (DB) absorption lines, or no spectral lines (DC) if the WD is cold enough to no longer excite atoms above their ground state ($\lesssim$ 11,000 K for DBs and $\lesssim$ 5,000 K for DAs). However, some WDs, especially cooler WDs, have been found with absorption features from heavier elements in their atmospheres. This has been interpreted as evidence for evolved planetary systems and surviving minor planets \citep{Debes2002, Jura2003, Zuckerman2007, Koester2014}.  

The first WD observed with metal pollution in its atmosphere was discovered in 1917 \citep{vanMaanen1917}. It was initially classified as an F-type star since its spectrum featured large amounts of Ca and Fe absorption. Six years later it was re-classified as a WD \citep{Luyten1923}, a newly coined stellar type, but the astrophysical implications of this went unrealized until nearly a century later. Heavy metals on WD surfaces are expected to have short diffusion timescales, on the order of days to at most a few million years, relative to the WD cooling time. Therefore, the presence of heavy metal lines in WD spectra indicate that the metals cannot be primordial, but must be recently accreted from rocky material \citep{Koester2006}. It was first conjectured that the metals were accreted from the interstellar medium (ISM) as older, and therefore cooler, WDs would have the chance to interact with dense ISM clouds frequently throughout their lifetimes, i.e., every 50 million years or so \citep{Dupuis1993}. At the time, this matched with observations of metal-polluted WDs because they became prevalent at temperatures cooler than 20,000 K, which equates to WDs that have cooling ages of approximately 65 million years old or older, depending on the white dwarf mass. However, \cite{Farihi2010} discovered that the accreted mass for a majority of the polluted WDs was equivalent to the mass of an asteroid or moon-like object, and could not possibly be explained through ISM accretion. Additionally, many metal-polluted WDs reside in regions where interactions with ISM clouds are rare, and occur within the same populations as non-polluted WDs, further suggesting that ISM accretion cannot be the main contributor to metal pollution \citep{Wesemael1982}. 


About 25 -- 50\% of all WDs below 20,000 K are polluted by heavy metals \citep{Zuckerman2010, Koester2014, Badenas2023}. Observations of intact planets around WDs \citep{Vanderburg2020, mullally2024} and transiting planetary debris \citep{Vanderburg2015, Vanderbosch2020, Vanderbosch2021} further lends the notion that planetary systems can persist through to the WD phase. Planets and planetesimals that survive the late-stage evolution of their host stars and enter the WD's Roche radius, possibly through orbital disruptions during mass loss in the asymptotic giant phase and dynamical interactions with other planets \citep{Maldonado2020}, become tidally disrupted and the debris is subsequently accreted onto the WD. Since the discovery of the first polluted WD, nearly 1400 WDs have been spectrally classified as metal-polluted (DZ, or DAZ, DZA, DBZ, DZB, etc. depending on the predominant lines in their spectra) \citep{Dufour2017}, with a majority only exhibiting Ca pollution. 
 
 In the next decade, we will begin to characterize the atmospheres of rocky exoplanets with current and next generation telescopes \citep{Wordsworth2022}. To complement these advances, it is key to probe the interior compositions of exoplanets. The chemical properties of exoplanets are commonly inferred using the bulk density, which is heavily model-dependent. However, WDs that have accreted planetary material provide a direct means to study planetary chemical structure as the innards of an exoplanet undergoing engulfment are on full display for us to study via spectroscopy. More specifically, the `heavily polluted' DZs with multiple metals present can be used to constrain exoplanet geological structure, formation, and destruction history \citep{Harrison2018, Xu2019, Hollands2021, Bonsor2023}. The presence of crust-, mantle-, and core-forming metal species in some DZs indicates that the accreted bodies were most likely chemically differentiated, similar to Earth \citep{Zuckerman2011, Wilson2015, Kawka2016, Swan2019}. Additionally, the fraction of WDs that have accreted core--mantle differentiated material can help us ascertain whether enrichment in short-lived radioactive nuclides like $^{26}$Al \citep{Jura2013, Curry2022}, which fuel the formation of Fe cores, is common to all star-forming regions \citep[e.g.][]{Young2016} or a rare feature of the Solar System \citep[e.g.][]{Gounelle2012,Lichtenberg2016}.  
Core or mantle-rich material could also indicate the accretion of a single large body \citep{Brouwers2022} or the signature of a collisionally evolved population of exo-asteroids \citep{Bonsor2020}. In order to probe geological processes, Ca, Mg, Fe, alongside Ni or Cr are required, and to identify crustal material, key elements Na, Li, or K are required \citep{Kaiser2021, Hollands2021}. Such elements have the strongest lines in cool DZs ($\lesssim$ 11,000 K), so expanding the number of DZ objects to use for abundance measurements of their planetary material is crucial. However, our portrait of exoplanet and planetesimal compositional diversity is incomplete as only a few dozen DZs \citep{Hollands2017, Hollands2018, Bonsor2023} have been found with enough metal species to place constraints on differentiation and formation conditions. As such, the emergence of large-scale telescopic surveys within the past couple of decades provides us the opportunity of finding an unprecedented number of heavily polluted DZs to use for studying the chemical diversity of exoplanets.

The most recent \gaia~data release (DR3) in 2022 \citep{GaiaDR3} revealed nearly 1.5 billion sources with full astrometry, an increase of nearly 136 million sources from the prior data release in 2019. The plethora of data from \gaia~has enabled advances in areas such as Galactic archaeology \citep{Poggio2018, Lucey2023, deason2024}, stellar evolution \citep{Jao2018, Fouesneau2023}, and measuring cosmic distances \citep{BailerJones2021, Riess2021, Ripepi2023}. \gaia~has also been pivotal in the field of WD astrophysics, increasing the number of WDs by nearly 100,000 after the second data release \citep{Gentile2019}, which contributed to such advancements as the discovery of the WD crystallization branch and cooling anomaly \citep{Tremblay2019, Cheng2019} and the development of a technique for finding new variable WDs using \gaia~excess photometric scatter \citep{Guidry2021}. The \gaia~observations consist of broadband (G), blue passband (BP), and red passband (RP) photometry that cover the wavelength ranges 330--1050 nm, 330--680 nm, and 640--1050 nm, respectively. In earlier data releases, BP and RP were simple photometric magnitudes. However, \gaia~DR3 integrated low resolution ($R=\lambda/\Delta\lambda \sim$ 70) blue and red grism spectra to create the BP and RP photometry, and released these spectra for 220 million sources. Despite the low resolution of the BP and RP (XP) spectra, they are invaluable for characterizing atmospheric parameters like effective temperature (\teff), surface gravity (\logg), metallicity, and line identification for vast quantities of stars \citep{Andrae2023, Zhang2023, Vincent2023}. 

In this paper, we combine the scale of the \gaia~XP spectra with the organizational power of an unsupervised machine learning method called Unsupervised Manifold Approximation and Projection (UMAP) \citep{UMAP2020} to find new members of scientifically intriguing populations of WDs such as the heavily polluted WDs. We use UMAP to organize the \gaia~XP spectra into distinct islands where each region corresponds to WDs with similar atmospheric qualities. The resulting archipelago of WDs allows us to pick out unique WD spectral groups such as the cool DZs for follow-up observations. Machine learning in conjunction with large-scale all-sky telescope surveys has unlocked massive potential for discovery, and we present one of the ways to exploit this union to find an unprecedented number of new cool DZs and a treasure trove of other types of WDs for upcoming follow-up studies. Using our new method, we have potentially increased the number of known cool DZs by nearly 300, with many of these likely harboring 5 or more metal species.


\section{Data}

\subsection{The Gentile-Fusillo WD Catalog}
\label{section:gentile}

To begin our journey of conjoining the \gaia~XP spectra and machine learning to find the polluted WDs, we must select a clean sample of WDs with \gaia~XP spectra. We draw our initial sample from the Gentile-Fusillo catalog of WD candidates in \gaia~DR3 \citep{Gentile2021}. This catalog was constructed by applying selection criteria based on \gaia~DR3 quality filters, absolute magnitude, and color to sources within the WD evolutionary sequence in the Hertzsprung-Russell (H-R) diagram. 
The \gaia~DR3 quality filters include cuts with respect to errors in proper motion, inconsistencies among the G, BP, and RP photometry (most likely due to nearby source contamination), and errors in the astrometric solutions of the sources. This results in a sample of $\sim$ 1.3 million sources in the WD region of the H-R diagram.

Since some objects, like quasars, tend to appear within the WD region in the H-R diagram, an additional filter is applied to the 1.3 million sources to single out the high-probability WDs and avoid contamination from non-WD objects. The WD probability parameter ($P_{WD}$) in the Gentile-Fusillo catalog provides a quantitative means of removing false positives from the catalog of WDs. 
Two density distributions were created for known WDs and contaminants in the \gaia~H-R diagram. These density distributions were then used to create a probability map, which is the ratio of the WD density distribution and the sum of the two distributions.
A 2D gaussian is assigned to all objects within the WD region, determined by errors in absolute magnitude and BP-RP color. The $P_{WD}$ parameter is then calculated by integrating the product of the Gaussian distribution for each source and the underlying probability map. After including only sources with $P_{WD} > 0.75$, the number of high-confidence WDs in the Gentile-Fusillo catalog is closer to 359,000, a nearly 40\% increase from the \gaia~DR2 WD catalog \citep{Gentile2019}. 

Other qualities such as \logg, \teff, and mass are provided for a majority of objects in the catalog. Many of these parameters are derived from a combination of the \gaia~photometry and astrometry, as well as atmospheric models for pure H, He, and mixed H/He. For the cool DZ WDs (\teff~$\lesssim$ 11,000 K), the calculations for the stellar parameters may be less reliable due to large flux suppression at bluer wavelengths ($\lesssim$ 4500 \AA) because of increases in opacity from metals like Ca and Fe. Therefore, some caution should be taken when using parameters that depend upon the atmospheric models in this catalog for the cool DZs. 

\subsection{The \gaia~XP Spectra}

 About 108,000 of the 359,000 WD candidates in the Gentile-Fusillo catalog also have \gaia~BP and RP (XP) spectra, which can be used for preliminary atmospheric classification. The \gaia~XP spectra are low resolution (R $\sim$ 70) grism spectra in the 330 -- 1050 nm wavelength range, where the BP range covers 330 -- 680 nm and the RP range covers 640 -- 1050 nm \citep{Carrasco2021}. Rather than storing the spectra as flux within corresponding wavelength bins, they are stored as coefficients corresponding to a linear combination of basis functions, specifically the Hermite functions. Hermite functions are the product of a Gaussian function and the normalized Hermite polynomials and can be represented as
 \begin{equation}
     H_G = H_{m}(x)~e^{-x^2}
     \label{eq:HG}
 \end{equation}
 where $H_G$ is the Hermite function, and $H_{m}(x)$ is the normalized Hermite polynomial with degree $m$, and $e^{-x^2}$ is the Gaussian function. The Hermite functions are orthonormal, meaning each of the basis functions independently contribute to the overall signal, and also converge to zero at large values of $x$. Much like the wavelength response of the \gaia~spectra, the Hermite functions approach zero for sufficiently long and short wavelengths \citep{Carrasco2021}.    

As a spectrum cannot be completely represented by a single basis function, a combination of multiple basis functions is used to capture all of the information encoded in the XP spectra. The \gaia~XP spectra are made up of 55 coefficients in each of the BP and RP bandpasses, summing to a total of 110 coefficients. 
A representation of the mean spectra in Hermite space is given in Eq. \ref{eq:hermite}, which is also described in Eq. 5 in \cite{Carrasco2021}.
\begin{equation}
h(u) = \sum_{n=0}^{N} c_n \cdot H_{G_n}(u)
\label{eq:hermite}
\end{equation}
$h(u)$ is the mean spectrum of the source with respect to a pseudo-wavelength $u$. $u$ does not have units of wavelength, but rather represents the sampling location in the focal plane of the BP and RP images. The $c_n$ represent the XP coefficients that correspond to the linear combination of Hermite functions $H_{G_n} (u)$ as shown in Eq. \ref{eq:HG}. The XP coefficients can be transformed to spectra if the appropriate set of basis functions are applied.

Fortunately, a \texttt{Python} package called \texttt{GaiaXPy} \footnote{\url{https://gaia-dpci.github.io/GaiaXPy-website/}} \citep{gaiaxpy} was created to convert the \gaia~XP coefficients to flux ($\mathrm{W~m^{-2}~nm^{-1}}$) vs. wavelength (nm) space. The pseudo-wavelengths are converted to units of wavelength and from Eq. \ref{eq:hermite}, a set of basis functions are applied to the coefficients to get the mean spectrum. 

  

\subsection{Sample Preparation}

Now that we have all WDs from the Gentile-Fusillo catalog with \gaia~XP spectra, we need to apply additional cuts to prepare our sample for categorization using UMAP. We want an astrometrically and photometrically clean sample to avoid ambiguous classifications. Thus, we choose quality cuts that minimize the impacts of systematic errors in the astrometry, flux, and parallax in our sample. At the same time, we do not want our cuts to be too strict such that a large percentage of WDs are excluded from our map. Listed below are the parameters we used for our quality cuts. 

\begin{enumerate}
    \item \texttt{Pwd} $> 0.9$
    
    This is the same $P_{WD}$ described in Section \ref{section:gentile}, but here we set the probability of an object being a WD to greater than 90\%. This is to further reduce possible contamination from other non-WD objects in our final classification sample.

    \item \texttt{phot\_g\_mean\_mag} $\leq 20$

    The apparent magnitude of the star in the photometric G band would affect the signal-to-noise of the spectrum. Even though it would be ideal to include only bright sources in our sample, most WDs are fainter than $\sim$ 18.5 mag, and we do not want our cleaned sample to be too exclusionary. If we make our cutoff at 19 or 19.5 mag, we would be excluding nearly 25 - 40\% of all WDs with \gaia~XP spectra. Therefore, we choose a more lenient magnitude cutoff of 20 mag.    



    \item \texttt{visibility\_periods\_used} $> 8$

    The number of visibility periods corresponds to the number of groups of observations separated by a minimum of four days. If the number of \gaia~transits for an object is large but there are not a substantial number of observations with long periods of time ($\geq$ 4 days) separating them, this may result in large errors in astrometric measurements like parallax that are unaccounted in the reported uncertainties. Requiring a minimum of 7--10 visibility periods ensures a lower likelihood of erroneous parallax and proper motion measurements \citep{quality_cuts}.

    \item \texttt{phot\_bp\_mean\_flux\_over\_error} $> 10$

    Because the BP and RP fluxes are summed over a relatively large aperture (3.5 $\times$ 2.1
  arcseconds$^2$), background flux contamination such as from nearby bright sources can have a large effect. Therefore, we place a 10\% limit on the photometric BP flux error. The {\it G} flux is much less susceptible to contamination from nearby sources since it comes from a narrow image of the source rather than integrated over a larger aperture like what is done for the BP-RP flux measurements \citep{Evans2018}. Therefore, we only include cuts in BP and RP flux errors.

    \item \texttt{phot\_rp\_mean\_flux\_over\_error} $> 10$

    This is the RP analog for the BP flux error described above.
\end{enumerate}

\begin{figure}
  \includegraphics[width=\linewidth]{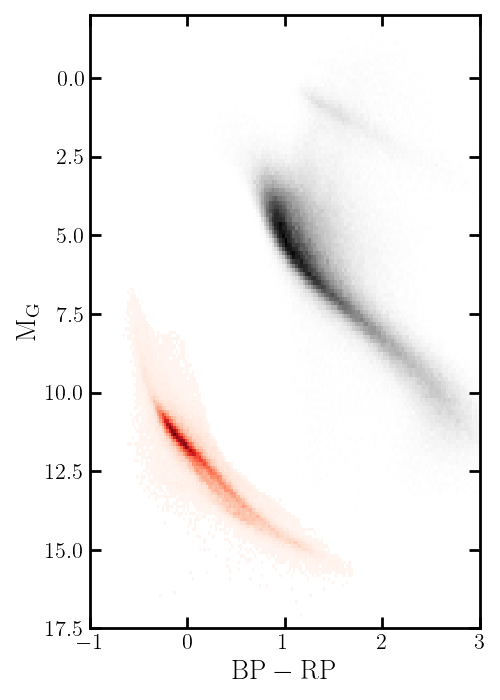}
  
\caption{About 1 million sources in \gaia~DR3 are plotted in grey along with our sample of 96,134 WDs in red in a color-magnitude diagram (CMD). The x- and y-axes correspond to BP-RP color and absolute G magnitude respectively. 
The main sequence stars are located within the large diagonal region in the CMD in grey. 
The WDs appear at $M_G \lesssim$ 7.5 and at BP-RP color $\lesssim$ 1.5. Our sample in red represents the WDs in \gaia~after applying our parallax, astrometry, and flux error cuts.}
\label{fig:HR}
\end{figure}

 Once we apply all of the quality control cuts on the initial sample of 108,000 probable WDs with XP spectra, we are left with a total of 96,134 WDs to use for classification. We plot our cleaned sample in a color-magnitude diagram (CMD) alongside 1 million randomly sampled stars in \gaia~DR3 shown in Fig. \ref{fig:HR}. The WDs are both under-luminous due to their small radii, and hot making them blue in color.     


\subsection{The Montreal White Dwarf Database}

To gain a better sense of how UMAP categorizes our final sample of 96,134 WDs, we require an existing database of known spectrally-classified WDs to compare within our unsupervised classification. The Montreal White Dwarf Database (MWDD) was created in 2016 as a go-to catalog for all known spectroscopically-classified WDs \citep[see][]{Dufour2017}. The spectroscopic data from 212 reference papers as well as from some online databases are compiled into tables that can be accessed online \footnote{\url{https://www.montrealwhitedwarfdatabase.org/tables-and-charts.html}}. 
To date, there are approximately 70,000 WDs in the MWDD categorized by atmospheric composition. In addition to spectral type, the MWDD table provides other characteristics such as variability, magnetism, and binarity. The current MWDD contains over 1370 DZs, but only a third of them have \gaia~XP spectra. We use these 425 DZs along with other WD classes in MWDD to interpret our categorization method. 

\section{Methods}
\label{section:methods}

\subsection{Categorizing \gaia~DR3 WDs with UMAP}

It has already been shown that the \gaia~XP spectra can be exploited through machine learning to label and infer stellar parameters for vast quantities of stars \citep{Andrae2023, Lucey2023, Sanders2023, Zhang2023}. A supervised machine learning technique trained on nearly
14,000 WDs with Sloan Digital Sky Survey (SDSS) spectra was applied to 100,000 WDs in the Gentile-Fusillo WD catalog with \gaia~XP spectra \citep{Vincent2023}. Supervised machine learning requires the use of a labeled dataset where each input feature  corresponds to an output label. The dataset trains the machine learning algorithm to recognize patterns and learn the
expected outcome for a given input. Once the algorithm is trained, it can be fed unlabeled data for which it will predict the output. In \cite{Vincent2023}, the SDSS training dataset included six spectral types for which to classify the WD \gaia~XP spectra: DA, DB, DC, DO, DQ, or DZ. Each WD is then assigned a probability for each spectral type. A disadvantage of this technique concerning the DZs is that the supervised machine learning tends to pick out the DZs with the strongest Ca II H\&K lines, which are some of the most prominent features in many DZ spectra, thereby excluding or misclassifying DZs with weaker Ca lines. 
\cite{Vincent2023} quotes 60\% precision, or true positive to false positive + true positive ratio, but $>$ 85\% recall, or true positive to true positive + false negative ratio, for the classification of DZs. Another supervised machine learning technique using Random Forest classification applied to \gaia~WDs within 100 pc with XP spectra \citep{GarciaZamora_2023} yielded 90\% precision for DZs, but a lower recall of $<$ 50\%. They assigned spectral types to 9,446 previously unclassified WDs, 132 of which were classified as DZ, using WDs in MWDD as a training set.    

We present a different, but complementary, method that utilizes the \gaia~XP spectra and an unsupervised machine learning tool called UMAP \footnote{\url{https://umap-learn.readthedocs.io/en/latest/}} \citep{UMAP2020} to organize our sample of 96,134 WDs into identifiable spectral groups. 
Using unsupervised machine learning to categorize low-resolution objects has already been shown to be successful. A similar manifold learning technique called t-Distributed Stochastic Neighbor Embedding (t-SNE) was implemented for finding metal-poor stars with low-resolution (R$\sim$750) spectra in the Hobby Eberly Telescope Dark Energy Experiment survey \citep{Hawkins2021}. Moreover, the UMAP method in conjunction with \gaia~XP spectra was previously utilized by \cite{Sanders2023} to separate C- and O-rich asymptotic giant branch stars. Unlike the supervised machine learning method \citep{Vincent2023}, unsupervised machine learning methods do not require a prior training dataset. Instead, they are free to interpret patterns and structures in the input features without instruction on the expected outcomes. Unsupervised machine learning is especially useful for detection of anomalous data points and clustering of data with similar attributes. 

 UMAP is a dimension reduction technique that takes in multi-dimensional data and collapses it into a 2D or 3D topological map that preserves the underlying structure of the data. The clusters and islands that appear in the map correspond to similarities in the input features interpreted by the algorithm. In this way, UMAP preserves both global and local structure of the data. For our case, each of the WDs in our sample contains 110 dimensions corresponding to the 110 XP coefficients. We normalize the coefficients by dividing them by the mean {\it G} flux to remove their brightness dependence. Otherwise, the {\it G} flux would inadvertently dominate the categorization method we intend to use to partition the WDs into unique spectral groups. We then input the normalized XP coefficients, which act as coordinates in the high-dimensional space, for all of the WDs into UMAP. The UMAP algorithm generates a surface, or manifold, on which the WDs are roughly uniformly distributed. We can then assign a radius to each WD that overlaps with neighboring WDs and reflects the local distances between them with respect to where they are located on the manifold. After the manifold is transformed into Euclidean space, or a flat geometry, the clustering of the WDs with similar features, as well as the distance between the clusters, is preserved. This output is the 2D map that we use for identifying the polluted WDs, among other unique WD groups.
 
 The two main hyper-parameters used in the construction of the map are \texttt{n\_neighbors} and \texttt{min\_dist}.

 \begin{enumerate}
     \item \texttt{n\_neighbors}

     This defines the number of neighboring data points that are included in the radius around a point of interest in the manifold. Higher values of \texttt{n\_neighbors} ($\gtrsim$ 100) would highlight the overarching structure of the data and forego the finer details, while lower values ($\lesssim$ 5) would only preserve the intricate connections of the data and lose the underlying global structure. Because we want to focus on emphasizing the sub-categorization of the data into unique spectral groups, but not so much that it becomes a detriment to global structure, we set \texttt{n\_neighbors} to 25.

     \item \texttt{min\_dist}

     This sets the minimum distance between points in the 2D map and affects its overall clumpiness. A \texttt{min\_dist} of 0 creates a very clustered map whereas 0.99 creates a very diffuse map. As we want to emphasize the congregation of points with similar spectral features, we set \texttt{min\_dist} to 0.05.
 \end{enumerate}
 
 The main advantages of UMAP over similar manifold learning techniques like t-SNE are computation time and global structure preservation. UMAP is not as sensitive to input data size or dimensions. On a modern laptop, a dataset with 100,000 points takes under 2 minutes to run in UMAP, but over 15 minutes in t-SNE \citep[see also Appendix C in][]{UMAP2020}. 
 Additionally, UMAP tends to preserve the global structure of the data more than t-SNE, so data within similar categories are more clearly co-located in different regions of the map. 
 The ability to cluster spectrally similar objects and to separate spectrally dissimilar objects result in distinct islands where objects like the cool DZs can easily be identified. 
 The UMAP yields a treasure trove of new DZs, especially those most likely contaminated with multiple metals, among other fruitful results with other unique WD groups. 



\section{Results}

\begin{figure}
\centering
  \includegraphics[width=\linewidth]{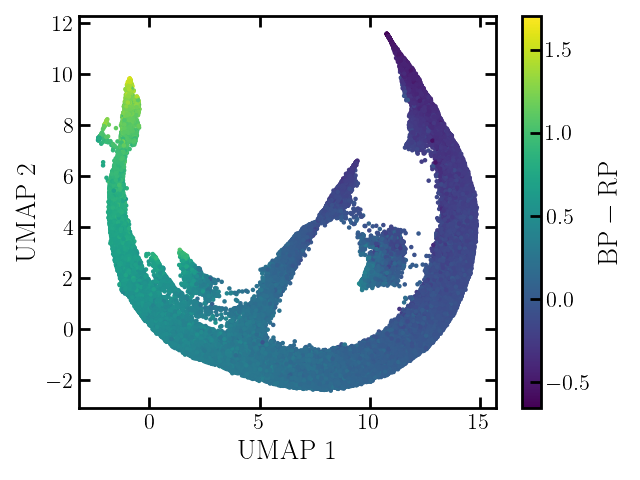}
  
\caption{The resulting map when passing the XP coefficients for 96,134 WDs through UMAP. The axes are unitless and only represent the 2D projection of the manifold as discussed in Section \ref{section:methods}. The map is colored by BP-RP color, where low (and/or negative) values of BP-RP represent bluer and therefore hotter WDs, while high values of BP-RP represent redder and cooler WDs.}
\label{fig:UMAP_color}
\end{figure}

Combining the advantages of large-sky-coverage spectroscopy with the \gaia~XP spectra and the clustering capabilities of UMAP unlocks new ways of discovering unique populations of WDs. We put all 110 of the BP and RP coefficients for the 96,134 input WDs in our sample through UMAP and end up with a 2D projection like that shown in Fig. \ref{fig:UMAP_color}.  

The map is organized into distinct regions that correspond to WD spectral groups with higher WD temperatures to the right and lower temperatures to the left. Using MWDD and the transformed \gaia~XP spectra, we can navigate through the UMAP terrain to find unique WDs such as the DAs, DBs, and cool DZs. We employ additional tools like RUWE and photometric variability to find other scientifically exciting, spectroscopically distinct groups such as the pulsating DA WDs (ZZ Cetis) and WD-M dwarf (WD-MD) binaries. The full table of the 96,134 WDs with UMAP coordinates will be available in machine readable format. A sample of our table is shown in Table \ref{table:UMAPWD}.

\begin{deluxetable*}{cccccc}
\tablehead{\colhead{WDJname} & \colhead{GaiaEDR3} & \colhead{RA (deg)} & \colhead{DEC (deg)} & \colhead{UMAP1} & \colhead{UMAP2}}
\startdata
WDJ065223.41+381228.84 & 944229799608599552 & 103.09746 & 38.207855 & 10.656315 & -1.3626201 \\
WDJ065223.58-111410.10 & 2953160246478762240 & 103.09827 & -11.236083 & 6.9203353 & 3.4263086\\
WDJ065225.02-204430.87 & 2932209288634116864 & 103.10416 & -20.741873 & 14.301202 & 5.6429334\\
WDJ065225.07+163918.83 & 3358160853443656704 & 103.10472 & 16.655012 & 1.3096164 & -0.4926479\\
WDJ065225.19+010130.11 & 3114030686095027328 & 103.10494 & 1.0251995 & 4.688952 & -0.6967064\\
... &... &... &... &... & ...
\enddata
\caption{The first five rows of our UMAP WD catalog with columns signifying the WDJ name, the \gaia~ERD3 source ID (GaiaEDR3), RA (in degrees), DEC (in degrees), and the UMAP coordinates (UMAP1 and UMAP2).}
\label{table:UMAPWD}
\end{deluxetable*}

\subsection{Cool DZs}
\label{cooldz}

\begin{figure}
\centering
  \includegraphics[width=\linewidth]{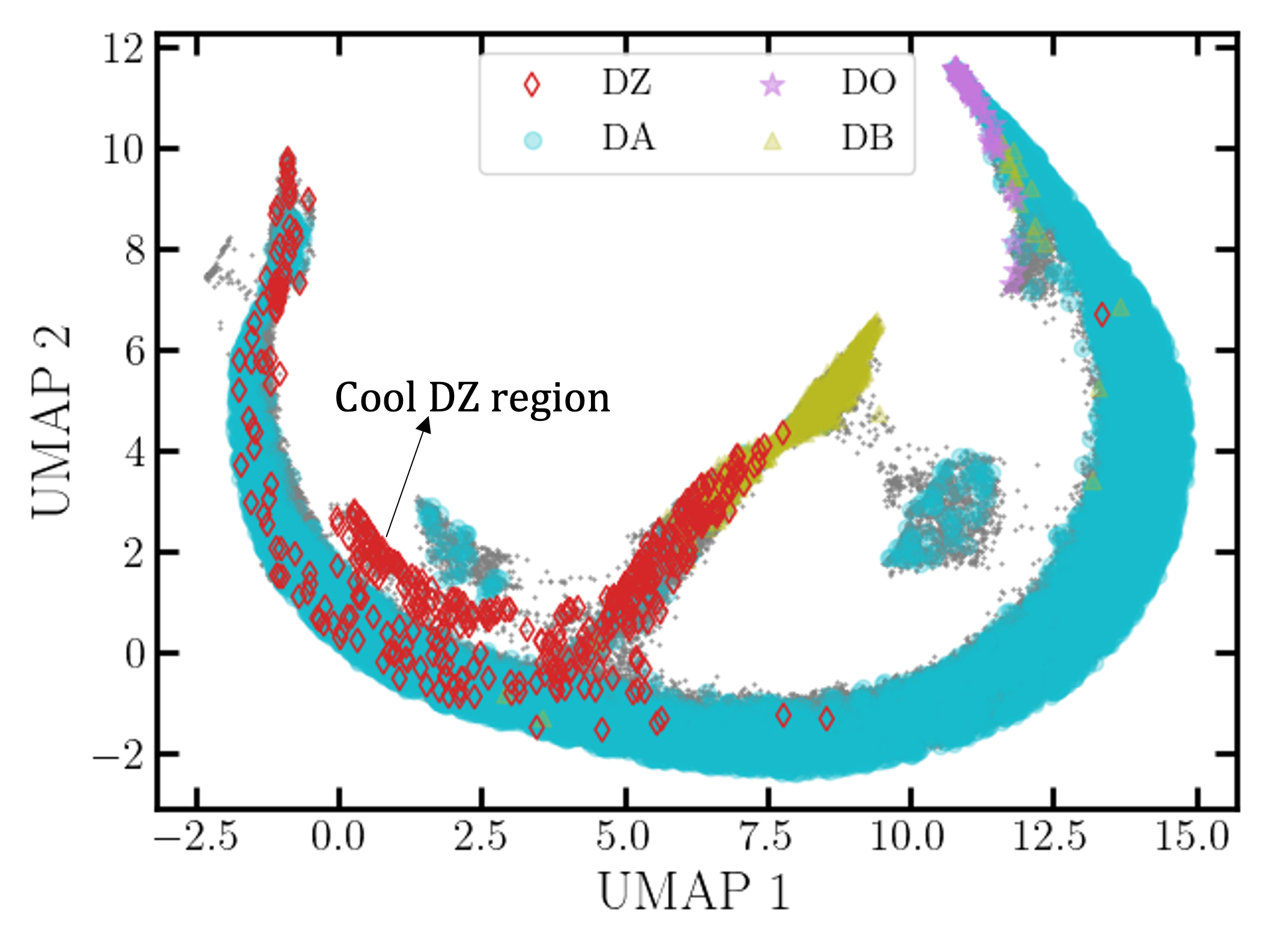}
  
\caption{The UMAP space with all DAs, DBs, DOs, and DZs from the MWDD overplotted with blue circles, gold triangles, purple stars, and red diamonds, respectively.}
\label{fig:MWDDDZ}
\end{figure}

To find the polluted WDs and other noteworthy WD groups, we must understand how UMAP assembled them in the 2D space. 
In Fig.~\ref{fig:MWDDDZ}, we plot known DAs, DBs, DOs, and DZs from MWDD in the UMAP space. The large horseshoe-shaped region that spans the entire BP-RP color spectrum of the map is dominated by DA WDs in MWDD. This is expected as 70\% of all WDs are thought to be DAs. The tip of the arm protruding from the center of the horseshoe almost exclusively contains DBs. The DBs are most prominent at temperatures below $\sim$ 40,000 K and above 11,000 K and exhibit He {\small I} absorption lines. Below 11,000 K, the He {\small I} lines begin to disappear as there is no longer enough energy to excite the electrons out of the ground state. The average temperature within the tip of the arm is around 17,000 K. Above $\sim$ 40,000 K, He-atmosphere WDs exhibit He {\small II} lines and are called the DOs, which tend to clump up at the rightmost tip of the horseshoe on the hottest end of the map. The DZs are mostly dispersed throughout the cooler half of the map with a large clump of DZs just above the DA horseshoe on the cooler end (\teff~$\sim$ 7,000~K) and spread throughout the bottom part of the DB arm (\teff~$\sim$ 11,000 K).   

\begin{figure}
\centering
\begin{minipage}{0.47\textwidth}
  \centering
  \includegraphics[width=\linewidth]{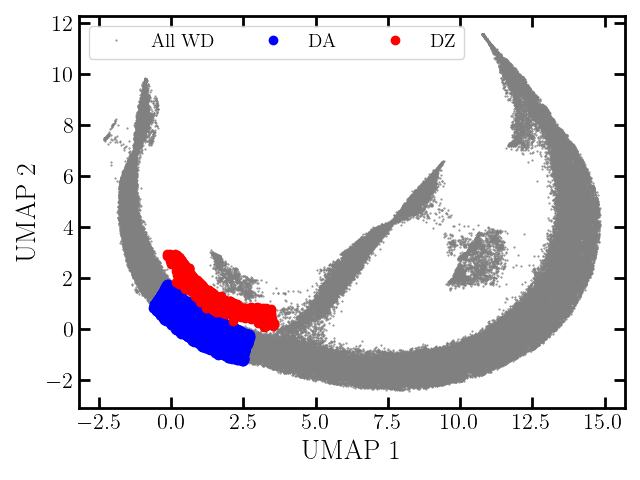}
\end{minipage}%
\begin{minipage}{0.47\textwidth}
  \centering
  \includegraphics[width=\linewidth]{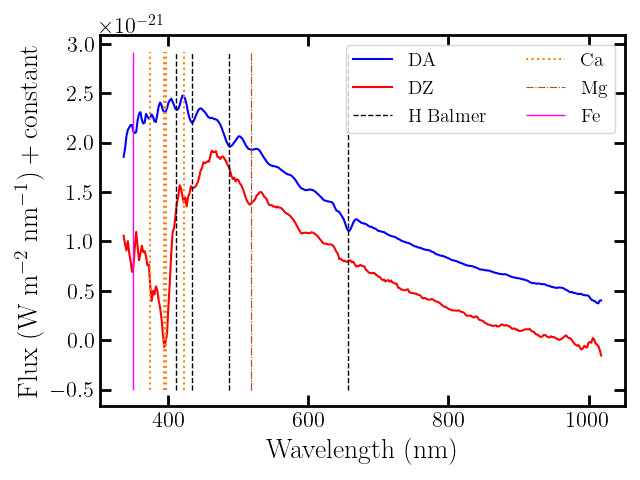}
\end{minipage}
\caption{Top: Locations in UMAP space of the cool DZ island (red) and the DA region of similar BP-RP color (blue). The DZ island is located just above the DA horseshoe. Bottom: The corresponding median-combined \gaia~spectra for both of the regions. The blue spectrum with H Balmer lines (656.3 nm, 486.1 nm, 434.0 nm, 410.2 nm), representing the median-combined spectra for 9273 WDs within the DA region, is shown on top. The red spectrum underneath, representing the median-combined spectra for the 475 WDs within the cool DZ island, exhibits significant absorption in the Ca {\small II} H\&K region (393.4 nm and 396.8 nm) as well as evidence of the Mg {\small I} {\it b} triplet around 520.0 nm and an Fe {\small I} line around 349.1 nm.}
\label{fig:poll_spectra}
\end{figure}

 Of particular note are the clump of DZs that appear within the cooler temperature region around the UMAP coordinates (2, 1) in Fig.~\ref{fig:MWDDDZ} and highlighted in the top panel of Fig.~\ref{fig:poll_spectra}.  
 At these low temperatures ($\sim$ 7000 K), WD atmospheres are very transparent, allowing certain metal species like Na and Li to be detectable \citep{Hollands2021}. We investigate this region for signs of significant metal pollution using \texttt{GaiaXPy} to transform the XP coefficients to a flux vs. wavelength spectrum in the bottom panel of Fig.~\ref{fig:poll_spectra}.
 

 The bottom panel of Fig.~\ref{fig:poll_spectra} demonstrates the difference between the appearance of the spectra in the DZ region and a ``typical" DA spectrum at approximately the same BP-RP color. The median-combined spectrum of the DA block is shown in blue and dashed blue lines indicate the expected locations of the H balmer line series (656.3 nm, 486.1 nm, 434.0 nm, 410.2 nm). This is compared to the median-combined spectrum of the DZ island shown in red with red dotted lines at strong Ca lines (373.7 nm, 393.4 nm, 396.8 nm, 422.6 nm), brown dot-dashed lines at the Mg {\small I} {\it b} triplet location (517.3 nm and 518.3 nm), and an Fe {\small I} line at 349.1 nm. Even though the Balmer lines in the DA spectrum are somewhat subdued because of the lower temperatures, they are still evident even at low resolution. In the red spectrum, there is very clear absorption from metal lines and an absence of clear He or H lines. This means we have unambiguously identified a region of cool DZs with significant metal pollution from at least three different metal species using UMAP.  

There are approximately 465 total candidates within this cool DZ region with the potential for harboring multiple metal species. Currently, there are a few dozen known DZs with five or more metal species present in their atmospheres \citep{Bonsor2023, Hollands2017, Hollands2018}. 375 of the 465 we have identified in the cool DZ island are not accounted for in MWDD, meaning that we could potentially increase the number of known heavily polluted WDs tenfold. About 80\% of the 375 are also classified as high-probability ($P >$ 65\%) DZs in \cite{Vincent2023}. The other 20\% show significantly less Ca absorption in their XP spectra compared to the high-probability DZs in the Vincent catalog, but still exhibit metal pollution. Similarly, 104 WDs found with the Random Forest method \citep{GarciaZamora_2023} show up in our polluted region, but only 75\% are classified as DZ. The other 25\% classified as DC or DA also exhibit metal pollution in their XP spectra, but are less dominated by Ca absorption. So while there is good correlation between the unsupervised and supervised methods, UMAP still uncovered several potential cool DZs that the supervised machine learning techniques overlooked. 



\subsection{RUWE and $V_\sigma$ in UMAP Space}
\label{section:variability}

Deducing the nature of stellar variability without access to detailed photometric light curves is often unfeasible. However, some clues can be extracted from metrics like the \gaia~renormalized unit weight error (RUWE) and from a variability metric based on the \gaia~photometric scatter of the star. Combining these within the UMAP terrain reveals variable classes of stars that would be difficult to identify through spectra alone.    

The \gaia~RUWE metric describes the amount of `wobble' around an astrometric solution for a given source and is tied to the likelihood of a star being a single star or in a multi-star system \citep{Belokurov2020, Penoyre2022}. RUWE is a re-normalization of the unit weight error (UWE), which is given by:  
\begin{equation}
    UWE = \sqrt{\frac{\chi^2}{N-p}}
\end{equation}
where {\em N} is the number of good observations of the source where the residuals agree with observational noise, $\chi^2$ is the goodness of fit of the source to a single star model, and {\em p} is the number of parameters used in the \gaia~astrometric model fit. 
It was shown in \cite{RUWE} that the UWE distributions and median values were dependent on $G$ magnitude and BP-RP color (see Figs. 1 \& 3 in \cite{RUWE} for reference). Therefore, an additional scaling to UWE with respect to color and magnitude was applied such that 
\begin{equation}
    RUWE = \frac{UWE}{UWE_{0}(G,C)}
\end{equation}
where $UWE_{0}(G,C)$ is a continuous function estimating a proxy for the mode of UWE in both color and magnitude \citep{RUWE}. The distribution of RUWE peaks around 1.0 for all stars with ``good" astrometry regardless of color or magnitude and can therefore be used as a gauge of possible binarity in stars with RUWE $>$ 1.25 \citep{Penoyre2022}. 

The variability of a source can also be determined through \gaia~parameters and, when paired with RUWE, can suggest whether the variability might be intrinsic or related to a companion. A metric for the photometric scatter of a star is the coefficient of variation for the {\em G} flux normalized by the number of observations of the star in the {\em G} band flux \citep{Guidry2021}. The photometric scatter can be approximated by 
\begin{equation}
    V_G = \frac{\sigma_G \sqrt{n}}{\bar{G}}
\label{eq:photscatter}
\end{equation}
where $\sigma_G$ is the error in {\em G} flux corresponding to the \gaia~parameter  \texttt{phot\_g\_mean\_flux\_error}, $n$ is the number of {\em G} band observations \texttt{phot\_g\_n\_obs}, and $\bar{G}$ is the mean {\em G} flux \texttt{phot\_G\_mean\_flux}. We bin $V_G$ in {\em G} magnitude and BP-RP color to minimize its color and magnitude dependence and apply a modified z-score to each of the bins. A z-score is a standardization method that assigns a value to each data point equivalent to the number of standard deviations, $\sigma$, away from the mean such that $z_i = (x_i-\bar{x})/s$. A modified z-score relies on the median of the sample, rather than the mean to minimize the influence of outliers on the z-score. In place of the sample standard deviation $s$, a median absolute deviation ($MAD$) is used instead. As outlined in section 3.3 in \cite{stats1993}, the modified z-score is given as 
\begin{equation}
    V_{\sigma}=0.6745 \times \frac{V_G - \widetilde{V_G}}{MAD}
\label{eq:vsigma}
\end{equation}
where $\widetilde{V_G}$ is the median $V_{G}$ from Eq. \ref{eq:photscatter} within the color-magnitude bin and $MAD$ is equivalent to $median\{|V_{G} - \widetilde{V_G}|\}$. 
The variability index for a given WD is represented by $V_{\sigma}$, where values $\leq$ 0.0 are considered non-variable. 

\begin{figure}
\centering
\begin{minipage}{0.47\textwidth}
  \centering
  \includegraphics[width=\linewidth]{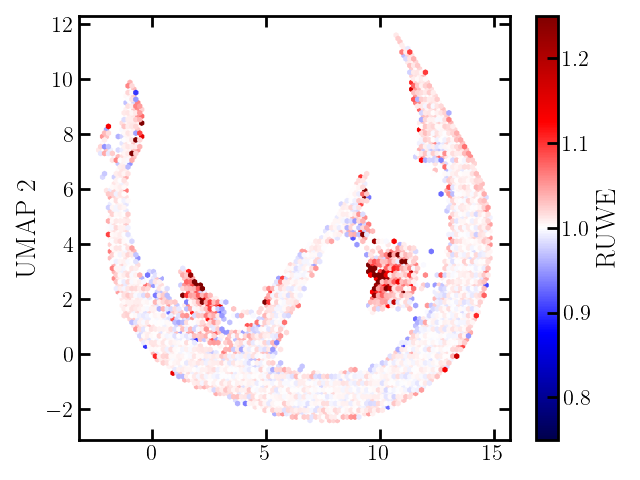}
  \label{fig:RUWE}
\end{minipage}%
\begin{minipage}{0.47\textwidth}
  \centering
  \includegraphics[width=\linewidth]{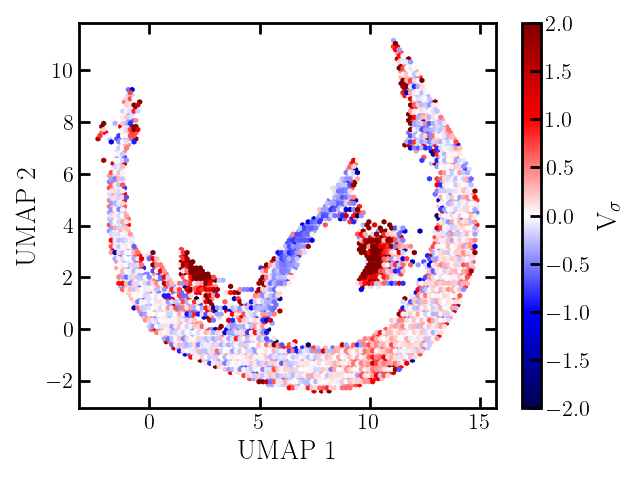}
  \label{fig:variability}
\end{minipage}
\caption{Top: The UMAP space with a color map according to RUWE. Objects with higher RUWE values have a higher likelihood of being in a multi-star system. Bottom: The UMAP space scaled with a color map according to photometric scatter, $V_\sigma$. Higher positive values of $V_\sigma$ indicate objects with high photometric scatter.}
\label{fig:RUWEandV}
\end{figure}

In Fig.~\ref{fig:RUWEandV} we plot RUWE and $V_\sigma$ as color maps in the UMAP space. The top panel of Fig. \ref{fig:RUWEandV} shows regions in the UMAP space where objects with high RUWE lie, indicating groups of WDs that have a higher likelihood of being in multi-star systems. The bottom panel of Fig. \ref{fig:RUWEandV} shows regions where the $V_\sigma$ is high, indicating the locations of WDs with excess photometric scatter in our map. We interpret WDs that have both high RUWE (detecable astrometric wobble) and $V_\sigma$ (detectable photometric variability) to likely have a stellar companion.
The regions where $V_\sigma$ is high but RUWE is low indicates photometric variability of a different nature. Interestingly, a strip of high $V_\sigma$ objects around the UMAP coordinates (10, -1) can be seen in the $V_\sigma$ color map, but not in the RUWE map. We identify these as the pulsating DAs.




\subsection{Even More Treasures: ZZ Cetis, DBs, and WD-MD Binaries}

\subsubsection{ZZ Cetis}
\label{sec:zzcetis}

In the bottom panel of Fig. \ref{fig:RUWEandV}, there is an evident increase in $V_{\sigma}$ in a strip within the DA horseshoe around the UMAP coordinates (10, -2) and (10, 0). However, this strip does not have high RUWE, as can be seen in the top panel of Fig. \ref{fig:RUWEandV}, signifying that the photometric variability is intrinsic in nature, rather than caused by a transiting companion. Some DA WDs exhibit pulsations reminiscent of the seismic waves we experience on Earth. 
As a WD cools, the free electrons in its atmosphere begin to recombine with the H ions, creating a partially ionized atmosphere \citep{McGraw1979}. The resulting increase in opacity  leads to the formation of a surface convection zone, and this convection zone can 
modulate the flux in such a way as to drive $g$-mode pulsations in these stars \citep{Robinson1982,Goldreich99b,Wu99}.
The temperature range for which pulsations in DAs are most prominent is the ZZ Ceti instability strip ($\sim$ 10,000 -- 13,000 K), named for the second star discovered of this class in the Cetus constellation. 
The ZZ Cetis can be used in asteroseismic studies to constrain quantities such as mass, chemical composition, rotation, and convection \citep{Winget2008,Fontaine08}.

\begin{figure}
\centering
  \includegraphics[width=\linewidth]{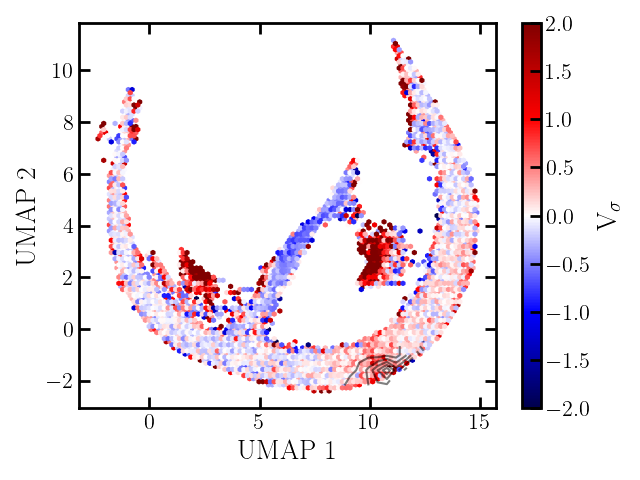}
  
\caption{This is the same figure as shown in the bottom panel of Fig.~\ref{fig:RUWEandV} with known ZZ Cetis in MWDD and in \cite{Vincent_ZZCeti} over-plotted with contour lines. The known ZZ Cetis tend to coincide with the high-variability region in the DA horseshoe.}
\label{fig:ZZCetis}
\end{figure}

Fig.~\ref{fig:ZZCetis} shows the variability map with contours (located near the UMAP coordinates (10, -2)) representing 200 known ZZ Cetis in \gaia~DR2 \citep{Vincent_ZZCeti} and single non-magnetic variable DAs within the temperature range of the instability strip in MWDD. This indicates where ZZ Cetis fall within the UMAP region. The location of the known ZZ Cetis matches with the increase in $V_\sigma$ in the DA horseshoe mentioned in Section \ref{section:variability}. Furthermore, this increase in variability does not correlate with an increase in RUWE, meaning that the origin of variability is unlikely due to binarity.  

Most known ZZ Cetis have masses around 0.6 \solarmass, but there are some rare massive ($\gtrsim$ 1 \solarmass) ZZ Cetis which likely have crystallizing cores. 
Just like any phase transition from a liquid to a solid, core crystallization leads 
to the release of latent heat. This additional energy source causes a delay in the WD cooling time, as evidenced by pile-ups in the WD luminosity function and in the CMD \citep{Winget2009, Tremblay2019}. For WDs $\gtrsim$ 1 \solarmass, the crystallization temperature coincides with the temperature range of the ZZ Ceti instability strip which allows us to probe their crystalline interiors with asteroseismology \citep{Metcalfe2004}. The physics of crystallization and its effects on WD interiors are still not well understood, and asteroseismology can potentially aid in this. However, we have very little pulsational data on the massive ZZ Cetis thus far.

\begin{figure}
\centering
  \includegraphics[width=\linewidth]{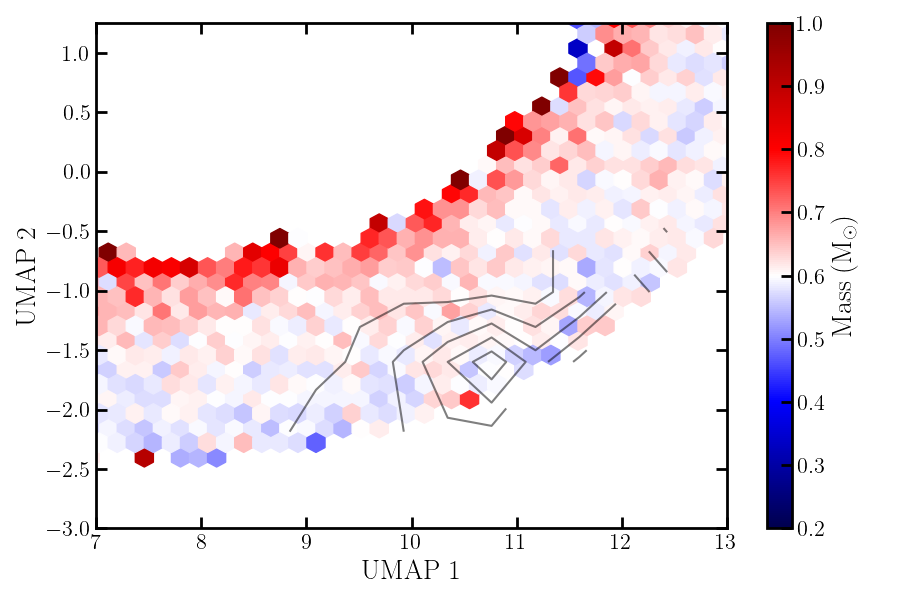}
  
\caption{Zoom-in of the ZZ Ceti region in the UMAP space, but with the color gradient mapping mass instead of $V_\sigma$. The contour lines from Fig.~\ref{fig:ZZCetis} are still over-plotted to indicate the location of the ZZ Cetis. Higher-mass ZZ Cetis are expected to appear near the inner boundary of the DA horseshoe.}
\label{fig:massiveZZ}
\end{figure}

In Fig.~\ref{fig:massiveZZ} we plot a zoom-in of the ZZ Ceti instability strip in the UMAP space with the color gradient according to mass. The masses come from the Gentile-Fusillo catalog and are calculated using mass-radius relations for C/O-core WDs with thick H atmospheres \citep{Bedard2020}. The higher mass WDs tend to clump up near the inner boundary of the DA horseshoe and could potentially reveal many new massive pulsator candidates. Currently, there are at least 6 known pulsators with masses greater than 1 \solarmass~\citep{Kanaan1998, Hermes2013, Curd2017, Vincent_ZZCeti, Kilic2023}. The UMAP candidate selection method reveals over 20 candidates with masses $\gtrsim$ 1 \solarmass~and \teff~within the temperature range where pulsations should be detectable. All candidates have $V_{\sigma} > 2$.  

A typical selection method for ZZ Cetis is to apply \teff~and photometric variability cuts to the region in the CMD where the ZZ Cetis are located. This often results in a sample that is contaminated by non-pulsating and non-DA objects. For example, \cite{Vincent_ZZCeti} used the CMD and \teff~cuts to select ZZ Cetis in \gaia~DR2, and they claim a $\sim$ 25\% contamination rate of non-DAs based on the amount of known non-DAs they removed from their initial selection. Additionally, applying photometric variability cuts does not exclude variable, non-ZZ Ceti objects such as cataclysmic variables, highly magnetic WDs, or even transiting debris WDs. In this way, selecting ZZ Ceti candidates using the UMAP method provides the advantage of removing many of these spectral contaminants present in the CMD method.

\subsubsection{DB WDs}

DOs and DBs are He-atmosphere WDs that feature two different types of He absorption. Hot He-rich WDs exhibit singly ionized (one electron is stripped) He {\small II} lines and are labeled as DOs whereas He-rich WDs below 40,000 K exhibit neutral He {\small I} lines and are labeled as DBs. 

\begin{figure}
\centering
\begin{minipage}{0.47\textwidth}
  \centering
  \includegraphics[width=\linewidth]{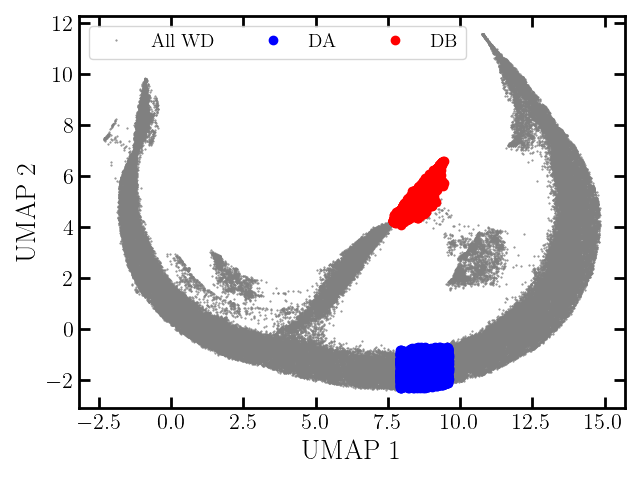}
\end{minipage}%
\begin{minipage}{0.47\textwidth}
  \centering
  \includegraphics[width=\linewidth]{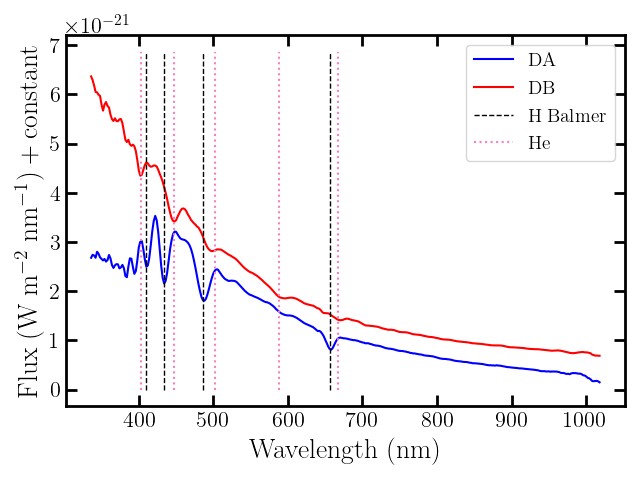}
\end{minipage}
\caption{Top: Locations in UMAP space of the DB tip (red) and the DA region of similar BP-RP color (blue). The DB tip coincides with the region where MWDD DBs dominate as shown in Fig.~\ref{fig:MWDDDZ}. Bottom: The corresponding transformed \gaia~spectra for both of the regions. The H Balmer series in blue dashed lines (656.3 nm, 486.1 nm, 434.0 nm, and 410.2 nm) can clearly be seen in the DA spectrum in blue. The He {\small I} lines (402.6 nm, 447.1 nm, 501.6 nm, 587.6 nm, 667.8 nm) shown with red dotted lines are visible in the DB spectrum in red placed above the DA spectrum.}
\label{fig:DB_spectra}
\end{figure}

One mystery in regard to the He-atmosphere WDs is the so-called DB gap, marked by a near absence of DBs in the 30,000 K $\leq$ \teff~$\leq$ 45,000 K range \citep{Koester2015, Torres2023}. After entering a quiescent stage of evolution, a small amount of H present in a DO atmosphere could float to the surface, making the WD appear as a hot DA. It would later be diluted by the onset of He convection around 30,000 K, therefore explaining the existence of the DB gap. However, because there are still DBs within the gap, albeit a small amount, the settling out of H may not occur in all DOs as some of them may have very little to no H in their atmospheres to begin with \citep{Koester2015}.   

With our UMAP technique, we have discovered a region of practically pure DB WDs. This DB region can be easily seen in Fig.~\ref{fig:MWDDDZ} and is shown more clearly with the red region in the top panel of Fig.~\ref{fig:DB_spectra} along with the corresponding median-combined XP spectra in the bottom panel. The DB spectrum in red is compared with a DA spectrum of similar BP-RP color in blue. The He {\small I} lines are evident in the DB spectrum.

There are 3694 WDs within the region outlined in the top panel of Fig.~\ref{fig:DB_spectra},
537 of which are classified as DBs in MWDD and 13 as DAs. This means that there potentially could be 3400 previously undiscovered DB WDs. 132 of the WDs included in the DB tip have temperatures between 30,000 and 40,000 K, according to the He-atmosphere photometric fits from the Gentile-Fusillo catalog. Approximately 70 DBs and DBAs (mixed He and H atmospheres) within the DB gap have been discovered thus far \citep{Eisenstein2006, Kleinman2013}, but our classification method could potentially double the number of WDs within this temperature range. 

\subsubsection{Cool WD-MD Binaries}

There are some WDs that are found in binary systems with low-mass M dwarf (MD) stars. MDs are the most common type of main-sequence (MS) stars and usually have masses below 0.6 \solarmass. 75\% of WD-MD binaries have long orbital periods, on the order of 100 days, and the other 25\% are in close binaries with orbital periods on the order of hours \citep{RebassaMansergas2013}. Sufficiently close binaries can undergo mass transfer from the MD to the WD and are potential progenitors for type Ia supernovae \citep{Heller2009}. Even though a large fraction of WD-MD binaries are believed to contain a cool WD companion or a dominant MD companion, they are often excluded from WD-MD binary studies \citep{RebassaMansergas2013}. This is because most spectroscopic WD-MS binary surveys only include observations in the optical wavelengths and not the infrared (IR), thus coverage of WD-MS binaries with \teff~$\lesssim$ 10,000 K is sparse \citep{Heller2009}.   

\begin{figure}
\centering
\begin{minipage}{0.47\textwidth}
  \centering
  \includegraphics[width=\linewidth]{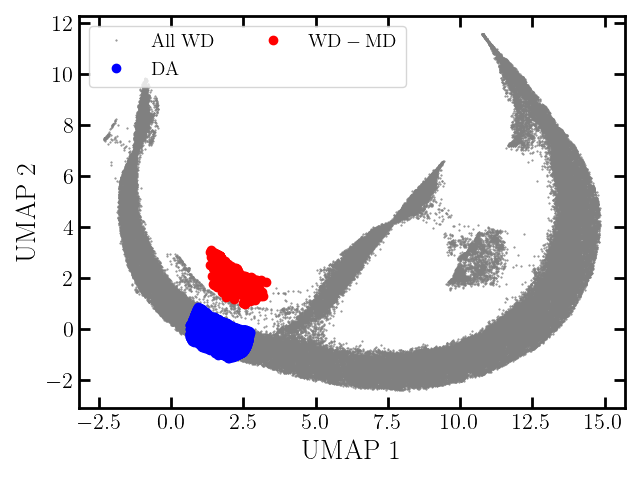}
\end{minipage}%
\begin{minipage}{0.47\textwidth}
  \centering
  \includegraphics[width=\linewidth]{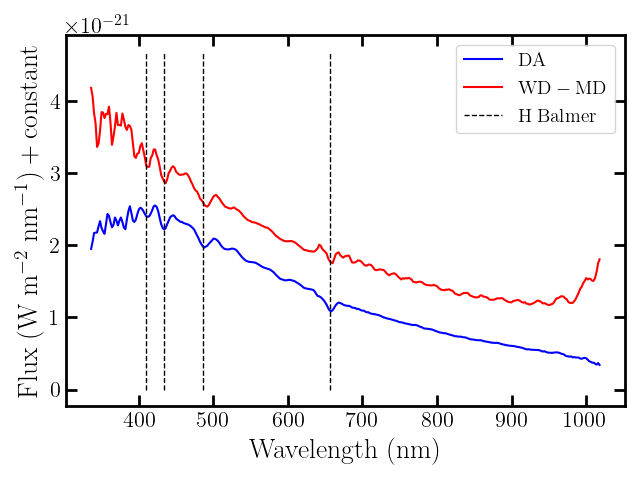}
\end{minipage}
\caption{Top: Locations in UMAP space of the WD-MD island (red) and the DA region of similar BP-RP color (blue). The WD-MD island is located above the DA horseshoe and the cool DZ island shown in Fig. \ref{fig:poll_spectra}. Bottom: The corresponding median-combined \gaia~spectra for both of the regions. The blue spectrum with H Balmer lines (656.3 nm, 486.1 nm, 434.0 nm, 410.2 nm) is shown on the bottom and corresponds to the blue DA block. H balmer lines are also present in the red spectrum above, but at the long-wavelength end, there is a noticeable flux excess.}
\label{fig:WDMD_spectra}
\end{figure}

Using the RUWE indicator and $V_\sigma$, we can locate the areas in the map that most likely are variable due to transiting binary companions. In Fig.~\ref{fig:RUWEandV}, the region corresponding to the WD-MD binaries in the top panel of Fig.~\ref{fig:WDMD_spectra} has both high $V_\sigma$ and RUWE values, signifying that the variability is most likely due to a passing companion.

We have identified 1096 objects within the cool WD-MD binaries region located above the DZ strip and the DA horseshoe in the top panel of Fig. \ref{fig:WDMD_spectra}. Six of these are already categorized in the MWDD as WD-MS binaries, and 39 as DAs. The average \teff~using the values from H atmospheric models in the Gentile-Fusillo catalog is around 7000 K. The bottom panel of Fig. \ref{fig:WDMD_spectra} shows the median-combined spectrum of the 1096 objects in red compared with the combined spectrum of the DA block of similar BP-RP color in blue. The H Balmer lines are present in both sets of spectra, but the red spectrum shows an uptick in flux at longer wavelengths, suggesting the presence of a cooler companion. The candidates within this region with IR excess may help to bridge the gap in data for cool WD-MD binaries. 





\section{Discussion}

 \begin{figure*}
\centering
  \includegraphics[width=\textwidth]{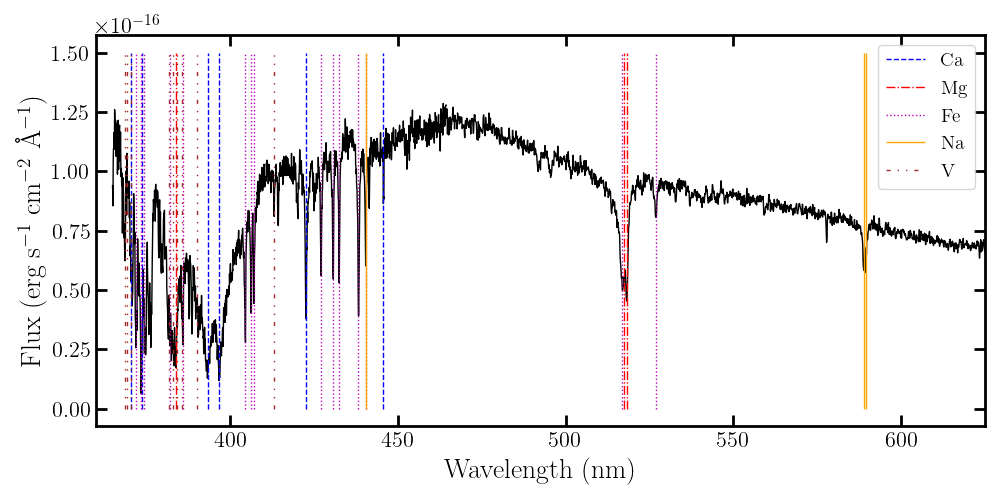}
  
\caption{A sample spectrum from our HET campaign with at least 5 metal species (Ca, Mg, Fe, Na, V) present. The Ca lines are shown with blue dashed lines, Mg with red dot-dashed lines, Fe with magenta dotted lines, Na with yellow solid lines, and V with brown dash-dot-dotted lines. The signal-to-noise per pixel for this spectrum is around 30.}
\label{fig:hetspectrum}
\end{figure*}
 
 Through the power of unsupervised machine learning combined with \gaia~XP spectra, we have discovered a multitude of new candidates within a variety of WD groups. Though we have identified an isolated island of 465 cool DZs, the other DZs within our sample of 96,134 WDs are not as distinctly separated from other spectral types. 
 As can be seen in Fig. \ref{fig:MWDDDZ}, the DZs in MWDD are spread throughout the map on the cool end. Many of the DZs cannot be distinguished from other spectral types like the DBs and DAs, because the metal lines were not strong enough at the resolution of the XP spectra to warrant a separation from the dominant element (H or He) in their atmospheres. About 21\% of DZs in MWDD in our \gaia~XP sample are within the cool DZ island (discussed in Section \ref{cooldz}), and the rest are spread throughout a higher temperature region contaminated with other spectral classes and in the DA horseshoe, not within a distinguishable island like the cool DZs. So we recognize that not all DZs with \gaia~XP spectra will be found in differentiable islands compared to WDs of other spectral types. 
 
 On the other hand, the more defined regions, like the cool DZ island, in the UMAP space have members that are spectrally similar, as evidenced by the median-combined \gaia~XP spectra and by the small amount of contamination from other members of differing spectral classification in MWDD. 
 For example, in the cool DZ island, there are no other spectral types that show up within it. To further demonstrate the high precision of our method, we have embarked on a higher-resolution spectroscopic campaign of the cool DZ island (Kao et al., in prep.). The 375 polluted candidates that we have identified using the UMAP method are currently being verified using the Hobby-Eberly Telescope (HET) at McDonald Observatory \citep{HET1998} using the LRS2 instrument \citep{LRS2} and the Very Large Telescope (VLT) at the European Southern Observatory (ESO) \citep{Albrecht1998} using the X-SHOOTER instrument \citep{Vernet2011}. So far, we have a 99\% detection rate for polluted WDs with multiple metal lines in their atmospheres. A sample spectrum for one of the confirmed candidates (\gaia~DR3 233493113910461952) is shown in Fig. \ref{fig:hetspectrum}. There are five predominant metal species within its spectrum: Ca, Mg, Fe, Na, and V. A follow-up paper will detail the results of our spectroscopic survey.


\section{Summary}

We have developed a novel method for discovering unique WDs in \gaia~using an unsupervised machine learning technique called UMAP to group the \gaia~XP coefficients in distinct regions within a 2D map. This enables the selection of spectrally similar objects with high precision for the purpose of finding new members of groups like the cool DZs that would be difficult to find via photometric methods and that are not available in other spectroscopic surveys. Many spectroscopic surveys lack broad magnitude coverage, so WDs, which are intrinsically faint, are often overlooked. While the \gaia~XP spectra do not permit detailed spectroscopic analysis due to their low resolution (R $\sim$ 70), they prove advantageous for categorizing large quantities of WDs by their dominant atmospheric qualities. An added benefit of the UMAP visualization of the XP spectra is mapping by variability, allowing us to pick out variable groups like the ZZ Cetis and WD-MD binaries. We have discovered 375 new polluted WDs with three or more metals present in their atmospheres which we are confirming with spectroscopic surveys that have yielded a 99\% success rate. 

\section{Acknowledgements}

This work has made use of data from the European Space Agency (ESA) mission
{\it Gaia} (\url{https://www.cosmos.esa.int/gaia}), processed by the {\it Gaia}
Data Processing and Analysis Consortium (DPAC,
\url{https://www.cosmos.esa.int/web/gaia/dpac/consortium}). Funding for the DPAC
has been provided by national institutions, in particular the institutions
participating in the {\it Gaia} Multilateral Agreement.

This project was developed in part at the 2023 Gaia XPloration, hosted by the Institute of Astronomy, Cambridge University.

The follow-up observations were obtained with the Hobby–Eberly Telescope, operated by McDonald Observatory on behalf of the University of Texas at Austin, Pennsylvania State University, Ludwig-Maximillians-Universität München, and Georg-August-Universität, Göttingen. The HET is named in honor of its principal benefactors, William P. Hobby and Robert E. Eberly. 

The Low Resolution Spectrograph 2 (LRS2) was developed and funded by the University of Texas at Austin McDonald Observatory and Department of Astronomy, and by Pennsylvania State University. We thank the Leibniz-Institut fur Astrophysik Potsdam (AIP) and the Institut fur Astrophysik Goettingen (IAG) for their contributions to the construction of the integral field units.

The Texas Advanced Computing Center (TACC) at the University of Texas at Austin provided
high-performance computing, visualization, and storage resources that have contributed to the
results reported within this paper.

D.E.W, M.H.M., B.H.D., K.A.H, and M.L.K. acknowledge support from the Wootton Center for Astrophysical Plasma Properties, a U.S. Department of Energy NNSA Stewardship Science Academic Alliance Center of Excellence supported under award numbers DE-NA0003843 and DE-NA0004149, from the United States Department of Energy under grant DE-SC0010623.
J.L.S. acknowledges support from the Royal Society (URF\textbackslash R1\textbackslash191555). AB acknowledges the support of a Royal Society University Research Fellowship, URF\textbackslash R1\textbackslash 211421.
M.H.M. acknowledges support from the NASA ADAP program under grant 80NSSC20K0455.

KH acknowledges Vasily Belokurov, JJ Hermes, and Nicola Gentile Fusillo for insightful conversations that led to this work. KH acknowledges support from the National Science Foundation
grant AST-2108736. This work was performed in part at the Simons Foundation Flatiron Institute’s Center for Computational Astrophysics during KH’s tenure as an IDEA Fellow.

\bibliography{bibliography,newrefs}{}
\bibliographystyle{aasjournal}

\end{document}